\documentclass[namedreferences]{SolarPhysics}
\usepackage[optionalrh,solaenum,natbib]{spr-sola-addons} 
\usepackage{graphicx}                    
\usepackage{color}                       
\usepackage{url,bm}

\graphicspath{{./fig/}{./png/}}

\newcommand{\EQ}{\begin{equation}}
\newcommand{\EN}{\end{equation}}
\newcommand{\EQA}{\begin{eqnarray}}
\newcommand{\ENA}{\end{eqnarray}}

\newcommand{\Fig}[1]{Figure~\ref{#1}}

\newcommand{\bra}[1]{\langle #1\rangle}

\newcommand{\meanrho}{\overline{\rho}}

{}
{}
{}

{}
{}
{}
{}
{}
{}
{}
{}
{}
\newcommand{\meanBB}{\overline{\mbox{\boldmath $B$}}{}}{}
{}
{}
{}
{}
{}
{}
{}
{}
{}
{}
\newcommand{\meanUU}{\overline{\bm{U}}}

{}
{}
{}

\newcommand{\meanB}{\overline{B}}

\newcommand{\meanU}{\overline{U}}

\newcommand{\meanJ}{\overline{J}}

{}

{}
{}

\newcommand{\uu}{\mbox{\boldmath $u$} {}}
\newcommand{\UU}{\mbox{\boldmath $U$} {}}

\def\bb{\bm{b}}
\newcommand{\BB}{\mbox{\boldmath $B$} {}}

\newcommand{\JJ}{\mbox{\boldmath $J$} {}}

\newcommand{\AAA}{\mbox{\boldmath $A$} {}}

\newcommand{\ff}{\mbox{\boldmath $f$} {}}

\newcommand{\grav}{\mbox{\boldmath $g$} {}}
\newcommand{\nab}{\mbox{\boldmath $\nabla$} {}}

\newcommand{\SSSS}{\mbox{\boldmath ${\sf S}$} {}}

\newcommand{\DD}{{\rm D} {}}

\def\Pm{\mbox{\rm Pr}_M}
\def\Rm{\mbox{\rm Re}_M}

\def\Rey{\mbox{\rm Re}}

\def\cs{c_{\rm s}}
\def\qpz{q_{\rm p0}}
\def\qp{q_{\rm p}}
\def\betap{\beta_{\rm p}}

\def\betastar{\beta_{\star}}
\def\betacrit{\beta_{\rm crit}}
\def\Peff{{\cal P}_{\rm eff}}

\def\qs{q_{\rm s}}

\def\kf{k_{\rm f}}

\def\Brms{B_{\rm rms}}

\def\urms{u_{\rm rms}}

\def\mut{\mu_{\rm t}}
\def\nut{\nu_{\rm t}}

\def\etat{\eta_{\it t}}
\def\etatz{\eta_{\rm t0}}

\def\mut{\mu_{\rm t}}

\def\Beq{B_{\rm eq}}
\def\Beqz{B_{\rm eq0}}
\def\tautd{\tau_{\rm td}}
\def\tauto{\tau_{\rm to}}

\def\half{{\textstyle{1\over2}}}

\def\onethird{{\textstyle{1\over3}}}

\newcommand{\kG}{\,{\rm kG}}

\newcommand{\yapj}[3]{ #1, {ApJ,} {#2}, #3}

\newcommand{\yapjl}[3]{ #1, {ApJ,} {#2}, #3}

\newcommand{\yan}[3]{ #1, {Astron.\ Nachr.,} {#2}, #3}

\newcommand{\yana}[3]{ #1, {A\&A,} {#2}, #3}

\newcommand{\yjfm}[3]{ #1, {J.\ Fluid Mech.,} {#2}, #3}

\newcommand{\ypf}[3]{ #1, {Phys.\ Fluids,} {#2}, #3}

\newcommand{\ypp}[3]{ #1, {Phys.\ Plasmas,} {#2}, #3}

\newcommand{\yal}[3]{ #1, {Astron.\ Lett.,} {#2}, #3}
\newcommand{\yjetp}[3]{ #1, {Sov.\ Phys.\ JETP,} {#2}, #3}

\newcommand{\ysci}[3]{ #1, {Science,} {#2}, #3}
\newcommand{\ysph}[3]{ #1, {Solar Phys.,} {#2}, #3}

\newcommand{\ypre}[3]{ #1, {Phys.\ Rev.\ E,} {#2}, #3}

\newcommand{\ybook}[3]{ #1, {#2} (#3)}

\newcommand{\sapj}[1]{ #1, {ApJ}, submitted}

\newcommand{\pan}[1]{ #1, {Astron.\ Nachr.}, in press}

\newcommand{\smn}[1]{ #1, {MNRAS}, submitted}
\begin{document}

\begin{article}

\begin{opening}

\title{Spontaneous formation of flux concentrations in a stratified layer}

%
\author{Koen Kemel$^{1,2}$\sep
Axel Brandenburg$^{1,2}$\sep
Nathan Kleeorin$^{3,1}$\sep
Dhrubaditya Mitra$^1$\sep
Igor Rogachevskii$^{3,1}$
}
%
\runningauthor{Kemel {\it et al.}}
\runningtitle{Spontaneous formation of flux concentrations}
%
  \institute{$^{1}$ Nordita, AlbaNova University Center,
    Roslagstullsbacken 23, SE-10691 Stockholm, Sweden,
    email: brandenb@nordita.org \\
   $^{2}$ Department of Astronomy, Stockholm University, SE-10691
   Stockholm, Sweden\\
   $^{3}$ Department of Mechanical Engineering, Ben-Gurion University of the Negev, POB 653, Beer-Sheva 84105, Israel\\
 }
\begin{abstract}
The negative effective magnetic pressure instability discovered
recently in direct numerical simulations (DNS) may play a crucial
role in the formation of sunspots and active regions
in the Sun and stars.
This instability is caused by a negative contribution
of turbulence to
the effective mean Lorentz force (the sum of turbulent and
non-turbulent contributions)
and results in formation of large-scale inhomogeneous
magnetic structures from initial uniform magnetic field.
Earlier investigations of this instability
in DNS of stably stratified, externally forced, isothermal
hydromagnetic turbulence in the regime of large plasma beta
are now extended into the regime of larger scale separation ratios
where the number of turbulent eddies in the computational domain is about 30.
Strong spontaneous formation of large-scale magnetic structures is
seen even without performing any spatial averaging.
These structures encompass many turbulent eddies.
The characteristic time of the instability
is comparable to the turbulent diffusion time, $L^2/\etat$,
where $\etat$ is the turbulent diffusivity and
$L$ is the scale of the domain.
DNS are used to confirm that the effective magnetic pressure does
indeed become negative
for magnetic field strengths below the equipartition
field.
The dependence of the effective magnetic pressure on the field
strength is characterized by fit parameters that seem to show
convergence for larger values of the magnetic Reynolds number.
\end{abstract}
\end{opening}

\section{Introduction}

The 11-year solar activity cycle manifests itself through the periodic
variation of the sunspot number.
Sunspots consist of vertical magnetic fields with a strength
of up to $3\kG$ \citep[see, e.g.][]{P79,O03}.
It is generally believed that these fields continue in a similarly
concentrated fashion also beneath the surface in the form of magnetic
flux tubes or fibers \citep{P82}.
It was thought that fibral magnetic fields constitute a lower energy
state and might therefore be preferred.
Tube-like magnetic fields were frequently seen in hydromagnetic turbulence
simulations \citep{Nor92,Bra95,Bra96}, but those tubes were similar to
the vortex tubes in hydrodynamic turbulence.
Those vortex tubes have typical diameters comparable to the viscous length.
Similarly, the aforementioned magnetic tubes have a thickness comparable
to the resistive length.
For the Sun, however, the resulting tube thickness would be too small
to be relevant for sunspots.

At the surface, strong magnetic flux concentrations also form in
regions of strong flow convergence,
but the size of these regions is too small
for sunspots, because sunspots are usually much bigger than a single
granular convection cell.
In fact, a typical sunspot can have a diameter of some 30 pressure
scale heights.
The tremendous size of sunspots has therefore been used as an argument
that they are not made
near the surface, but at much greater depth far away from the surface.
At the bottom of the convection zone the convection cells are big enough
and could in principle be responsible for producing much bigger flux
concentrations.
Magnetic flux tubes can also form through the action of shear
as shown by various simulations \citep{Cline,GK11}.
While shear is likely an important ingredient of the solar dynamo,
it remains unclear whether the resulting magnetic
tubes are really able to
produce sunspots as a result of them piercing through the solar surface
and, more importantly, whether one should consider them being tied to
the deep shear layers of the Sun.

In this paper we discuss an alternative scenario in which sunspots are
shallow  phenomena that are not anchored at the bottom of the convection zone.
Various mechanisms have been discussed.
Of particular interest here are mechanisms that are based
on the suppression of turbulence by magnetic fields.
In the mechanism of \cite{KM00} it is the suppression of the turbulent
heat flux while in the mechanism of \cite{RK07} it is
the suppression of the turbulent hydromagnetic pressure.
Both mechanisms lead to a linear large-scale
instability in a stratified medium.
However, these mechanisms may be of different importance in different layers.

In this study we focus mainly on the second mechanism, which has recently been studied in direct numerical simulations (DNS) as well as in mean-field calculations.
This mechanism is called the  negative effective magnetic pressure
instability (or NEMPI for short).
NEMPI is a convective type instability that is similar to
the interchange instability in plasmas \citep{T60,N61,P82}
and the magnetic buoyancy instability \citep{P66}.
However, the free energy in interchange and magnetic buoyancy instabilities is due to the gravitational field, while in NEMPI it is provided by the small-scale turbulence.
NEMPI is caused by the suppression of turbulent
hydromagnetic pressure by the mean magnetic field.
When fluid Reynolds numbers is larger than 1 and the mean magnetic field is less than  the equipartition magnetic fields,
the negative turbulent contribution to the mean
Lorentz force is enough large so that the
effective mean magnetic pressure (the sum of turbulent and non-turbulent contributions) become negative \citep{KRR90,RK07}.
This is the main reason for the excitation of the large-scale instability that results in formation
of large-scale inhomogeneous magnetic structures.

The effect of the suppression of the turbulent
heat flux has not yet been studied as extensively as NEMPI
since the original paper of \cite{KM00}, although \cite{KO06} have used
that model to study the decay of sunspots.
Moreover, there is now some evidence that the effects anticipated by
\cite{KM00} may have already been in operation in various
simulations of solar convection where the spontaneous formation
of pores has been seen \citep{Stein}.
Another example is the large-eddy simulation
of \cite{KKWM10} where an imposed vertical
large-scale magnetic field
gets concentrated into giant vortices.
This result is reminiscent of that of \cite{Tao98}, who found a
segregation into magnetized and nearly unmagnetized regions in stratified convection simulations.
However, it has not yet been possible to obtain the large-scale
magnetic structures resembling the sunspots, in models with a strong
imposed magnetic flux tube structure at the bottom of the domain
\citep[see, e.g.][]{Rem09a,Rem09b,Cheung,Rem11a,Rem11b}.

In the rest of this paper, we focus on NEMPI, which was first
found in mean-field calculations of a stratified layer
\citep{KMR96,RK07,BKR10,BKKR11}.
However, those results remained unconvincing until NEMPI was also
discovered in DNS \citep[][hereafter BKKMR]{BKKMR}.
It it therefore most appropriate to begin our discussion with the latter.

\section{The model}
\label{model}

Following the earlier work of BKKMR,
we solve the equations for the velocity $\UU$,
the magnetic vector potential $\AAA$, and the density $\rho$,
\begin{equation}
\rho{\DD\UU\over\DD t}=-\cs^2\nab\rho+\JJ\times\BB+\rho(\ff+\grav)
+\nab\cdot(2\nu\rho\SSSS),
\end{equation}
\begin{equation}
{\partial\AAA\over\partial t}=\UU\times\BB+\eta\nabla^2\AAA,
\end{equation}
\begin{equation}
{\partial\rho\over\partial t}=-\nab\cdot\rho\UU,
\end{equation}
where $\nu$ is the kinematic viscosity, $\eta$ is the magnetic diffusivity
due to Spitzer conductivity of the plasma,
$\BB=\BB_0+\nab\times\AAA$ is the magnetic field,
$\BB_0=(0,B_0,0)$ is the imposed uniform field,
$\JJ=\nab\times\BB/\mu_0$ is the current density,
$\mu_0$ is the vacuum permeability,
${\sf S}_{ij}=\half(U_{i,j}+U_{j,i})-\onethird\delta_{ij}\nab\cdot\UU$
is the traceless rate of strain tensor, and commas denote
partial differentiation.
The forcing function $\ff$ consists of random, white-in-time,
plane non-polarized waves with a certain average wavenumber.
The forcing strength is such that the turbulent rms velocity is approximately
independent of $z$ with $\urms=\bra{\uu^2}^{1/2}\approx0.1\,\cs$.
The gravitational acceleration $\grav=(0,0,-g)$ is chosen such that
$k_1 H_\rho=1$, which leads to a density contrast between 
bottom and top of $\exp(2\pi)\approx535$.
Here, $H_\rho=\cs^2/g$ is the density scale height.
We consider a domain of size $L_x\times L_y\times L_z$ in
Cartesian coordinates $(x,y,z)$, with periodic boundary conditions in
the $x$ and $y$ directions and stress-free perfectly conducting
boundaries at top and bottom ($z=\pm L_z/2$).
The volume-averaged density is therefore constant in time
and equal to its initial value, $\rho_0=\bra{\rho}$.

Our simulations are characterized by the scale separation ratio, $\kf/k_1$,
the fluid Reynolds number $\Rey\equiv\urms/\nu\kf$,
the magnetic Prandtl number $\Pm=\nu/\eta $ and
the magnetic Reynolds number $\Rm\equiv \Rey \, \Pm$.
Following earlier work \citep{BKKR11}, it is clear that NEMPI
is more effective for small values of $\Pm$, so here we choose
$\Pm = 0.5$ and $\Rm$ in the range $0.7$--$70$.
The magnetic field is expressed in units of the local
equipartition field strength near the top, $\Beq=
\sqrt{\mu_0\rho} \, \urms$, while
$B_0$ is specified in units of the volume
averaged value, $\Beqz=\sqrt{\mu_0\rho_0} \, \urms$.
In addition to visualizations of the actual magnetic field,
we also monitor $\Delta\meanB_y=\meanB_y-B_0$, where $\meanB_y$ is an
average over $y$ and a certain time interval $\Delta t$.
Time is expressed in eddy turnover times, $\tauto=(\urms\kf)^{-1}$.
For comparison, we also consider the turbulent-diffusive timescale,
$\tautd=(\etatz k_1^2)^{-1}$, where $\etatz=\urms/3\kf$ is
the estimated turbulent magnetic diffusivity.
Another diagnostic quantity is the rms magnetic field in the
Fourier mode of $k=k_1$, referred to as $B_1$,
which is here taken as an average over
$2\leq k_1z\leq3$, and is close to the top at $k_1 z=\pi$.
(Note that $B_1$ does not include the imposed field $B_0$ at $k=0$.)

The simulations are performed with the {\sc Pencil Code},%
\footnote{{\tt http://pencil-code.googlecode.com}}
which uses sixth-order explicit finite differences in space and a
third-order accurate time stepping method.
We use numerical resolutions of $128^3$ and $256^3$ mesh points
when $L_x=L_y=L_z$, and $1024\times128^2$ when $L_x=8L_y=8L_z$.
To capture mean-field effects on the slower turbulent-diffusive
timescale, which is $\tautd/\tauto=3\kf^2/k_1^2$ times smaller than
the dynamical timescale, we perform simulations for several thousand
turnover times.

\begin{figure}[t!]\begin{center}
\includegraphics[width=\columnwidth]{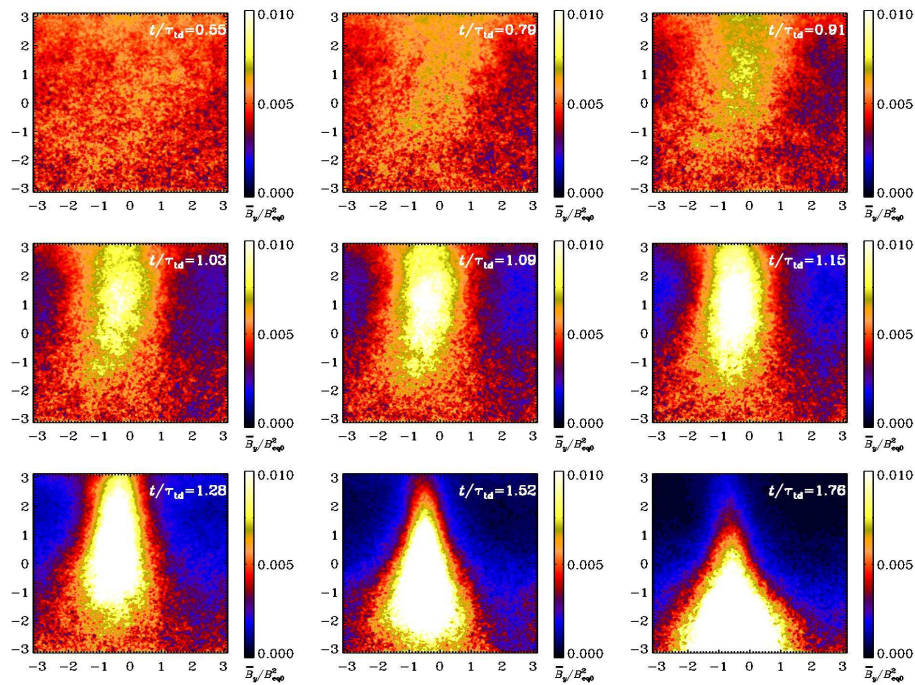}
\end{center}\caption[]{
Visualizations of $\meanB_y(x,z,t)$ for different times.
Time is indicated in turbulent diffusive times, $(\etatz k_1^2)^{-1}$, corresponding
to about 5000 turnover times, i.e., $t=5000/\urms\kf$.
$\Rm=18$ and $\Pm=0.5$.
}\label{Bm}\end{figure}

\section{Results}
\label{Results}

\begin{figure}[t!]\begin{center}
\includegraphics[width=\columnwidth]{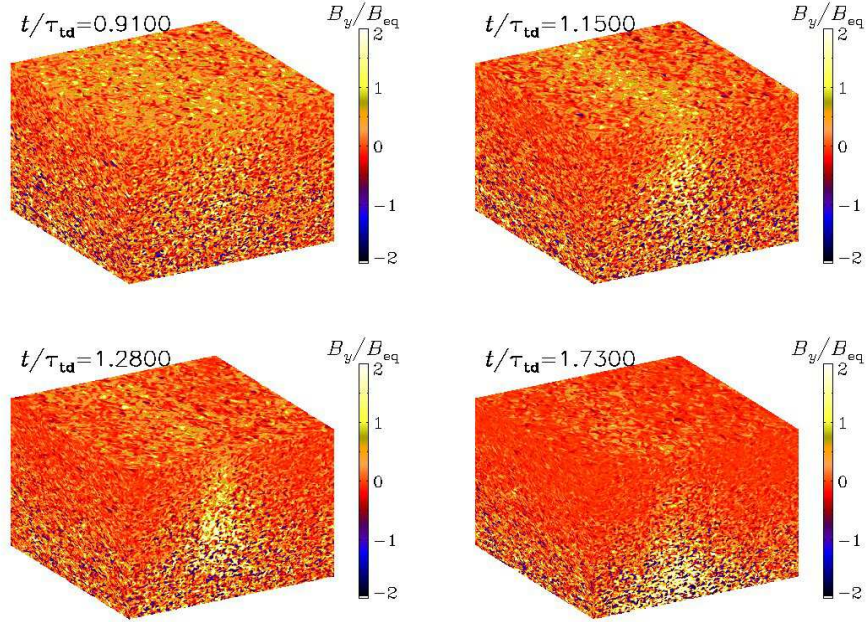}
\end{center}\caption[]{
Visualizations of $B_y$ on the periphery of the domain for different times,
indicated in turbulent-diffusive times, for the same run as in \Fig{Bm}.
}\label{B}\end{figure}

In simulations,
the clearest indication of a spontaneous flux concentration is seen
when the scale separation ratio is large.
In BKKMR, we only used $\kf/k_1=15$.
Here we consider calculations where this ratio is twice as large.
A useful diagnostic is the magnetic field averaged along the direction
of the imposed field, i.e., along the $y$ direction.
In particular, we shall be looking at the $y$ component of the field,
i.e., we look at $\bra{\Delta B_y}_y/\Beq$.
To see the effect even more clearly, we perform an additional time average
over about one hundred turnover times.
This average is then referred to as $\bra{\Delta B_y}_{yt}$.
In \Fig{Bm} we show $\bra{\Delta B_y}_{yt}/\Beq$ for $\kf/k_1=30$.
The other parameters are $\Rm=18$ and $\Pm=0.5$.
An inhomogeneous magnetic
structure forms first near the surface (at $t/\tautd=0.79$),
but then the structure propagates downward.
This is consistent with our interpretation that this is caused
by negative effective magnetic pressure operating on the scale
of many turbulent eddies.
Indeed, a local decrease of the effective magnetic pressure must be
compensated by an increase in gas pressure, which implies higher density,
so the structure becomes heavier and sinks in the nonlinear stage of NEMPI.
This is also seen in three-dimensional visualizations without averaging;
see \Fig{B} with the same parameters as in \Fig{Bm}.

\begin{figure}[t!]\begin{center}
\includegraphics[width=\columnwidth]{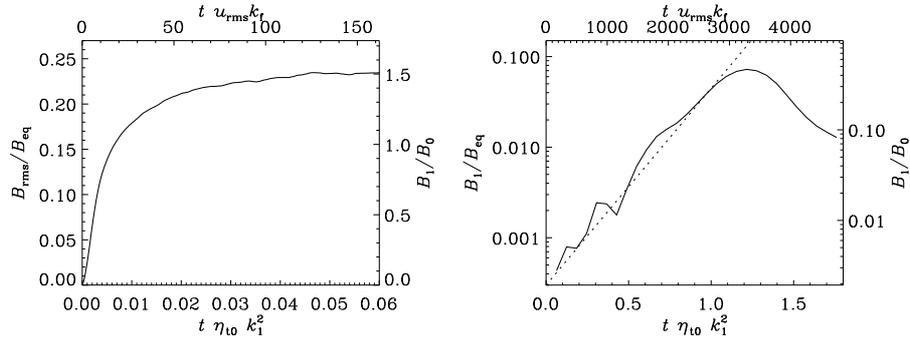}
\end{center}\caption[]{
Time dependence of $\Brms$ (left panel) and $B_1$ (right panel)
for the same run as in \Fig{Bm}, where
$\kf/k_1=30$ and $B_0/\Beqz=0.05$.
}\label{ppbymxztm2_I256k30c}\end{figure}

To confirm that NEMPI really is a large-scale instability,
we would expect to see an exponential growth phase.
This is shown in the right-hand panel of \Fig{ppbymxztm2_I256k30c},
where we show the growth of $B_1$ versus time.
We give time both in turbulent-diffusive times (lower abscissa)
as well as in eddy turnover times (upper abscissa).
We do indeed see that there is an exponential growth phase which
lasts for about one turbulent-diffusive time, i.e.,
the growth rate is comparable to $(\etatz k^2)^{-1}$, where
$\etat\approx\urms/3\kf$ is the expected turbulent magnetic diffusivity.
However, compared with the eddy turnover time, $(\urms\kf)^{-1}$,
the turbulent-diffusive time scale is $3(\kf/k_1)^2$ times slower.
This illustrates that NEMPI is indeed a very slow process compared
with, for example, the saturation of the overall rms magnetic field
(left-hand panel of \Fig{ppbymxztm2_I256k30c}).

While the rate of NEMPI may be too slow to explain the
relatively rapid appearance of sunspots on a time scale of days,
it would explain how one can produce magnetic structures on
length and time scales much bigger than the naturally occurring
scales in the upper layers of the Sun.
This discrepancy is exactly one of the reasons why one normally
places the formation of active regions at the bottom of the
convection zone \citep{Gol81}.
Here we see a clear example that this conclusion may be not correct.

\begin{figure}[t!]\begin{center}
\includegraphics[width=\columnwidth]{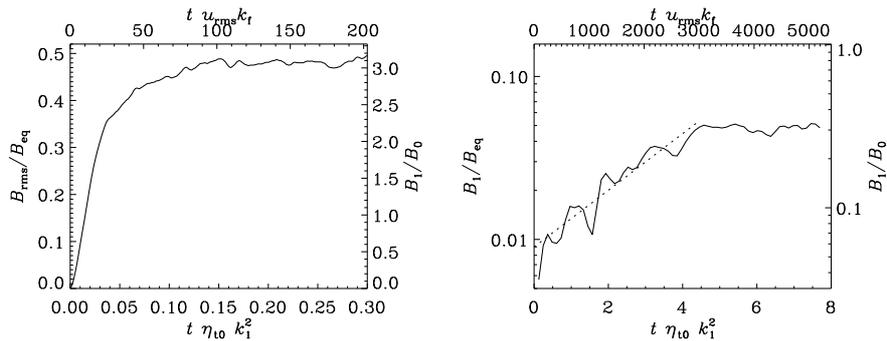}
\end{center}\caption[]{
Time dependence of $\Brms$ (left panel) and $B_1$ (right panel)
for a run used in BKKMR with $\kf/k_1=15$ and $B_0/\Beqz=0.05$.
}\label{ppbymxztm2_I128k15c}\end{figure}

Next, we turn to a simulation where the scale separation ratio,
$\kf/k_1$, is only half as big, i.e., $\kf/k_1=15$.
In that case we also see an exponential growth phase, but the
growth is slower (even in terms of turbulent-diffusive times),
lasts longer, and amplifies $B_1$ only by about
one order of magnitude; see \Fig{ppbymxztm2_I128k15c}.

\begin{figure}\begin{center}
\includegraphics[width=\columnwidth]{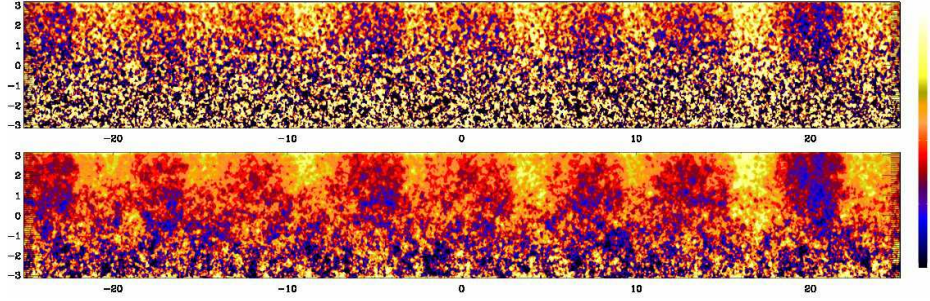}
\end{center}\caption[]{
Visualization of $\meanB_y(x,z)$ for an elongated box
($1024\times128^2$ mesh points) with $\Rm=36$
at a time during the statistically steady state.
The top panel shows the $y$ average $\bra{\Delta B_y}_y/\Beq$
at one time while the lower panel shows an additional time average
$\bra{\Delta B_y}_{yt}/\Beq$ covering about 80 turnover times.
}\label{pbymxz10}\end{figure}

\begin{figure}\begin{center}
\includegraphics[width=\columnwidth]{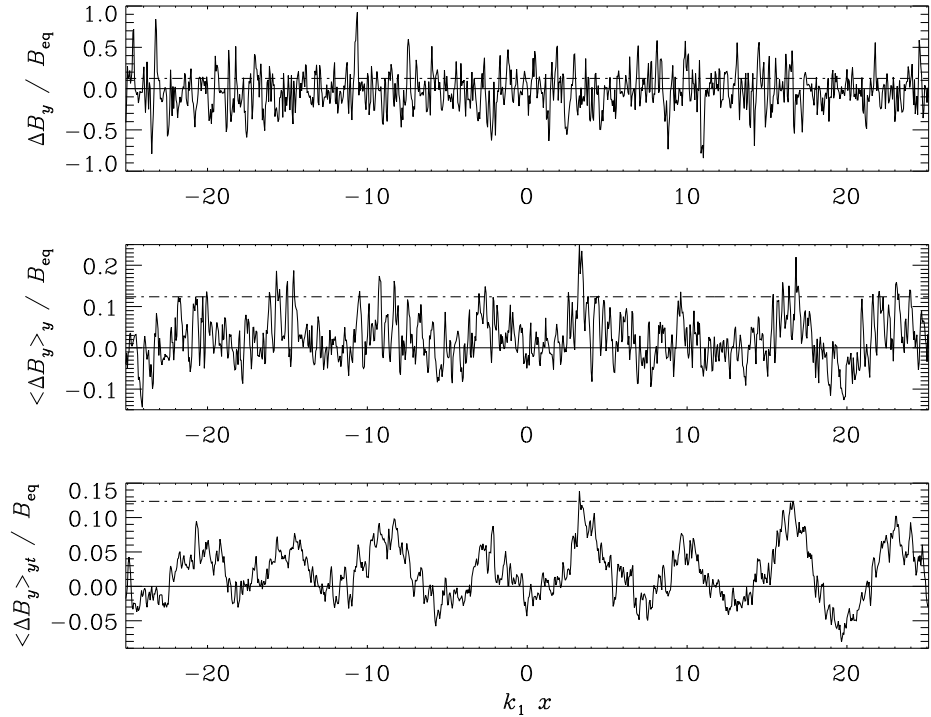}
\end{center}\caption[]{
$x$ dependence of the field for an elongated box
($1024\times128^2$ mesh points) with $\Rm=36$
at $k_1 z=2$ showing $\Delta B_y/\Beq$ at $y=0$
(top panel), its $y$ average $\bra{\Delta B_y}_y/\Beq$ (middle panel),
as well as an additional time average $\bra{\Delta B_y}_{yt}/\Beq$
(bottom panel) covering about 80 turnover times.
The dash-dotted line gives the level of the imposed field.
}\label{pbymxz10_slice}\end{figure}

Again, in agreement with related work involving the suppression
of turbulent heat flux, the most unstable mode has a horizontal
wavelength comparable to the vertical scale height of the layer;
see Fig.~2 of \cite{KM00}; hereafter referred to as KM.
This is also the case for NEMPI, as is shown in \Fig{pbymxz10}, where
we compare an instantaneous plot of $\bra{\Delta B_y}_{y}(x,z)$
with a time averaged one, $\bra{\Delta B_y}_{yt}$, making the
appearance of large-scale structures even more pronounced.
Next, in \Fig{pbymxz10_slice} we compare cross-sections of $\Delta B_y(x)$
(for fixed values of $y$, $z$, and $t$; top panel) with corresponding $y$
averages (middle panel) and $yt$ averages (bottom panel).
Here, $\Rm=36$, $\kf/k_1=15$, and $B_0/\Beqz=0.05$.
Without averaging, no clear magnetic structure is seen yet,
but the structures
become clearly more pronounced with $y$ and $t$ averaging.
Runs with similar parameters have been shown in BKKMR, for a computational
domain whose $x$ extent is $2\pi/k_1$ instead of $16\pi/k_1$.

The large-scale flux concentrations have an amplitude of only
$B_1 \approx0.1\Beq$
and are therefore not easily seen in single snapshots,
where the field reaches peak strengths comparable to $\Beq$.
Furthermore, as for any linear instability, the flux concentrations
form a repetitive pattern, and are in that sense similar to flux
concentrations seen in the calculations of KM that were
based on the magnetic suppression of the turbulent heat flux.
However, there are indications that at larger values of $\Rm$,
flux concentrations occur more rarely, which might be more
realistic in view of astrophysical applications.

\section{Quantifying the negative effective magnetic pressure effect}
\label{Formalism}

An important condition for the formation of structures by the mechanisms
of KM and \cite{RK07} is sufficient scale scale separation.
In other words, both the suppression of convective heat flux and the
suppression of total effective pressure have in common that they only work if a substantial number of eddies is involved in a turbulent structure under consideration.
If this is not the case, turbulent transport coefficients become
increasingly inefficient.
This is a natural feature of large-scale effects such as this one.
Here, we mean by turbulent transport coefficients any of the mean-field
coefficients that relate correlation functions of small-scale quantities
in terms of large-scale quantities.

Small- and large-scale quantities refer
here simply to a suitably defined averaging
procedure so that the velocity $\UU$, for example, can be split
into a mean (or large-scale) quantity $\meanUU$ and a fluctuation
$\uu=\UU-\meanUU$.
An important example of a turbulent transport coefficient is the
turbulent viscosity that emerges when relating the Reynolds stress
$\overline{u_i u_j}$ to the spatial derivatives of
the mean flow $\meanUU$ in the averaged
momentum equation,
\EQ
{\partial\over\partial t}\meanrho\meanU_i
=-{\partial\over\partial x_j}
\left(\meanrho\meanU_i\meanU_j+\meanrho\,\overline{u_i u_j}+...\right).
\label{nseq}
\EN
Here, $\meanrho$ is the average density, and correlations with density
fluctuations are neglected.
The simplest parameterization for $\overline{u_i u_j}$ is
\EQ
\overline{u_i u_j}=-\nut(\meanU_{i,j}+\meanU_{j,i})
-\mut\delta_{ij}\nab\cdot\meanUU ,
\EN
where $\nut$ is the turbulent viscosity
and $\mut$ is the turbulent bulk viscosity.
This relation is also known as the Boussinesq ansatz,
especially when contrasted with representations where the
$\Lambda$ effect is included, which is responsible for
producing differential rotation in the Sun \citep{Rue89}.
Here, however, we shall focus on magnetic effects.

In general, when there are magnetic fields,
the right-hand side of equation of motion~(\ref{nseq})
has to be replaced by the sum of Reynolds and Maxwell stresses
\EQ
\overline{\Pi}^{\rm f}_{ij}\equiv\meanrho\,\overline{u_i u_j}
-\overline{b_i b_j}/\mu_0+\half\overline{\bb^2}/\mu_0,
\EN
where the superscript f indicates contributions from the fluctuating field.
Furthermore, in the presence of a {\it mean} magnetic field, symmetry
arguments allow one to write down additional components, in particular
those proportional to $\delta_{ij}\meanBB^2$ and $\meanB_i\meanB_j$.

Expressing $\overline{\Pi}^{\rm f}_{ij}$ in terms of the mean field, the leading terms are
\EQ
\overline{\Pi}^{\rm f}_{ij}=\qs\meanB_i\meanB_j/\mu_0
-\half\qp\delta_{ij}\meanBB^2/\mu_0+...,
\label{A1}
\EN
where the dots indicate the presence of additional terms such as
the turbulent viscosity term mentioned earlier and terms that enter
when the stratification affect the anisotropy of the turbulence further.
Note in particular the definition of the signs of the terms involving
the functions $\qs(\meanBB)$ and $\qp(\meanBB)$.
This becomes clear when writing down the mean Maxwell stress resulting
from both mean and fluctuating fields, i.e.,
\EQ
-\meanB_i\meanB_j/\mu_0+\half\delta_{ij}\meanBB^2/\mu_0
+\overline{\Pi}^{\rm f}_{ij}
=-(1-\qs)\meanB_i\meanB_j/\mu_0+\half(1-\qp)\delta_{ij}\meanBB^2/\mu_0+...
\label{A2}
\EN
Thus, the signs are defined such that for positive $\qs$ and $\qp$ the
effects of magnetic stress and magnetic pressure are reduced and the
signs of the net effects may even change.
Equations~(\ref{A1}) and~(\ref{A2}) have been derived using the
spectral $\tau$ relaxation approach \citep{KRR90,KMR96,RK07} and the
renormalization procedure \citep{KR94}.

\begin{figure}[t!]\begin{center}
\includegraphics[width=\columnwidth]{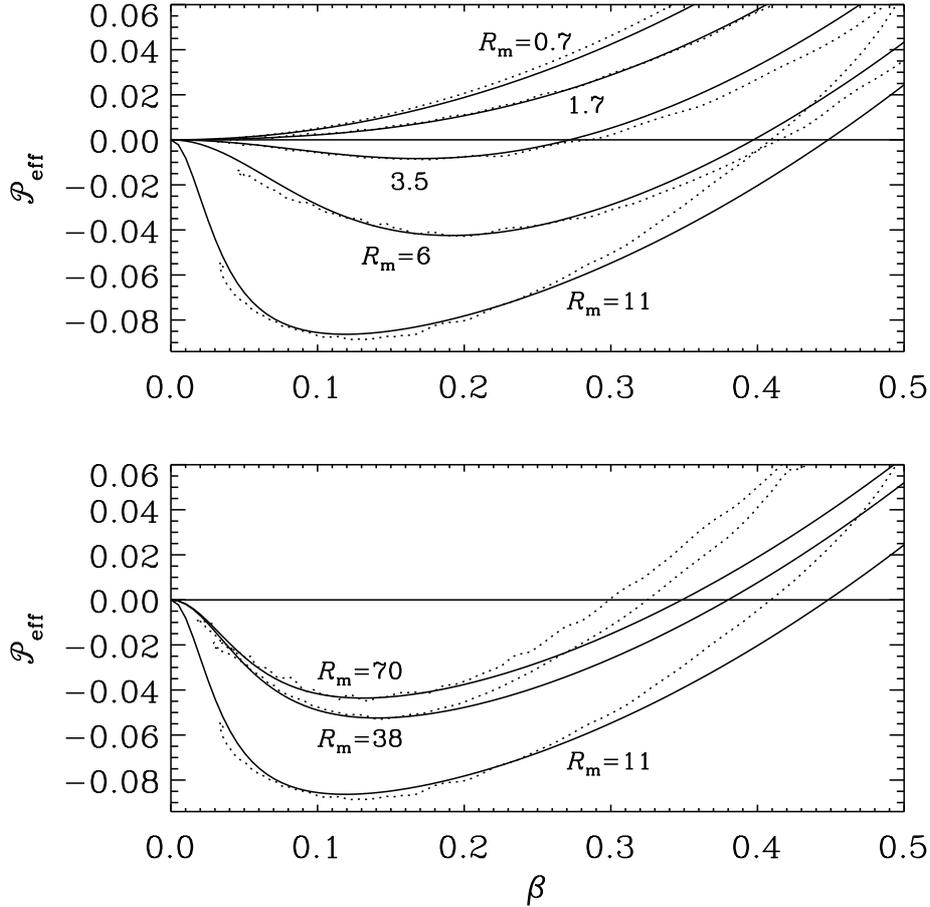}
\end{center}\caption[]{
Normalized effective magnetic pressure, $\Peff(\beta)$, for low (upper panel)
and higher (lower panel) values of $\Rm$.
The solid lines represent the fits to the data shown as dotted lines.
}\label{ppresseff}\end{figure}

A broad range of different DNS
in stratified turbulence \citep{BKKMR,BKKR11} or turbulent convection \citep{KBKMR11}
have now confirmed that $\qp$ is positive for $\Rm>1$, but $\qs$ is small and negative.
A positive value of $\qs$ (but with large error bars) was originally
reported for unstratified turbulence \citep{BKR10}.
Later, stratified simulations with isothermal stable stratification
\citep{BKKR11} and convectively unstable stratification \citep{KBKMR11}
show that it is small and negative.
Nevertheless, $\qp(\meanBB)$ is consistently larger than unity provided $\Rm>1$
and above unity while $\meanB/\Beq$ is below a certain critical value that is around 0.5.
This is shown in \Fig{ppresseff}, where we plot the effective magnetic
pressure,
\EQ
\Peff(\beta)=\half[1-\qp(\beta)] \, \beta^2\quad\mbox{versus}\quad
\beta\equiv|\meanBB|/\Beq
\EN
for different values of $\Rm$ using $\Pm=0.5$ and  $\kf/k_1=15$.
Note that the minimum of $\Peff(\beta)$ is deeper for the case
with $\Rm=11$ and becomes then shallower.

It should be noted that the coefficients for the fit formulae are
chosen such that the plot is most accurate for small values of $\beta$,
because this is what is most important for the onset of NEMPI.
Especially for larger values of $\Rm$ the fit formula does no longer
reproduce correctly the critical value $\beta=\betacrit$ for which $\Peff$
changes sign.

Note that $\betacrit$ is well below unity.
This implies that it is probably not possible to produce flux
concentrations stronger than half the equipartition field strength.
So, making sunspots with this mechanism alone is maybe unlikely,
and other effects such as that of KM may be needed.
It is possible that this mechanisms works preferentially in the
uppermost layers, provided enough flux has already been accumulated.
This may then be achieved with NEMPI which also works in somewhat
lower layers.

\begin{figure}[t!]\begin{center}
\includegraphics[width=\columnwidth]{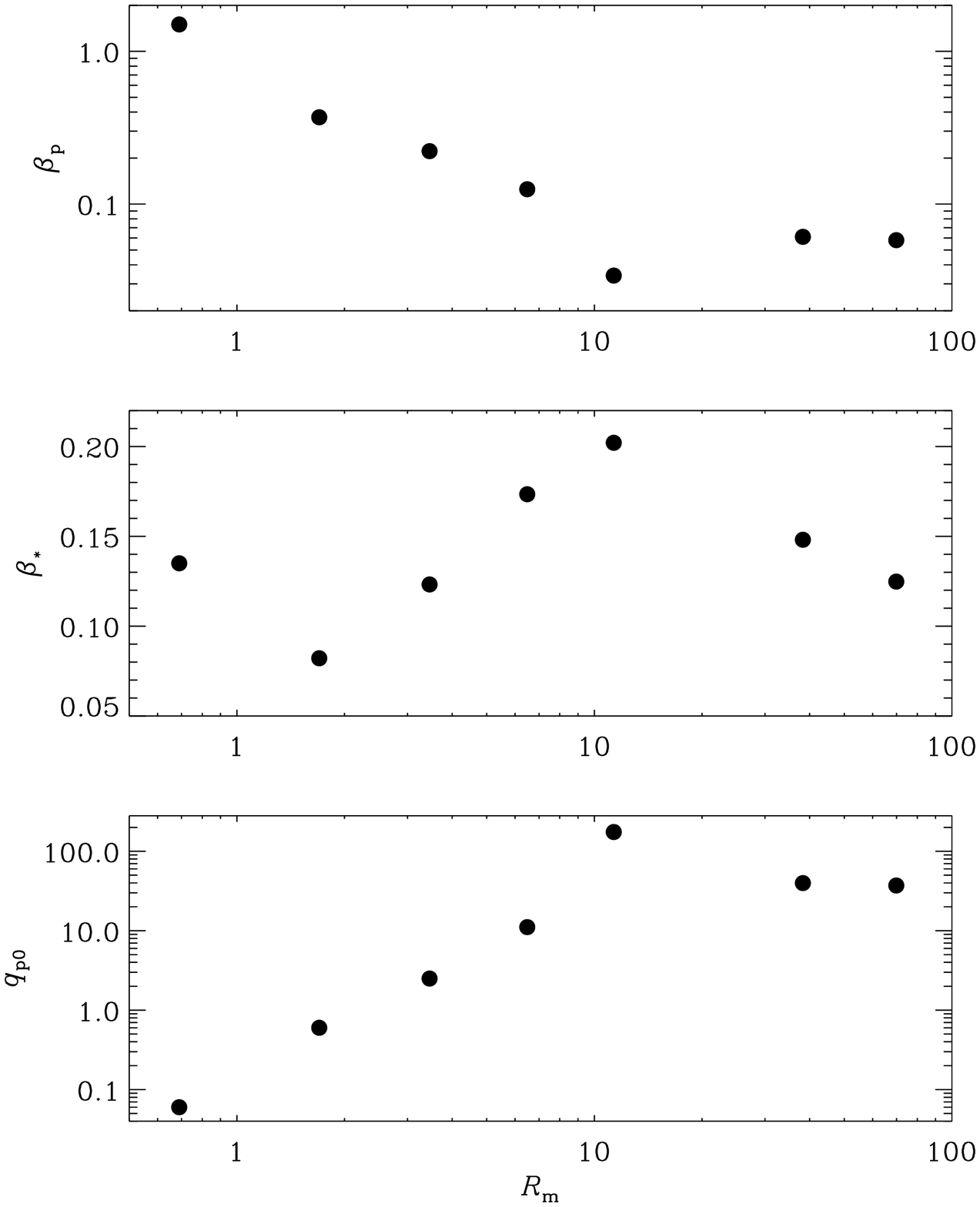}
\end{center}\caption[]{
$\Rm$ dependence of $\betap$, $\betastar$, and $\qpz$ for $\Pm=0.5$
and $\kf/k_1=15$.
}\label{pfitR}\end{figure}

To compare the resulting functions $\Peff(\beta)$ in a systematic
fashion for different parameters, we use the fit formula \citep{KBKR11}
\begin{equation}
\qp(\beta)={\qpz\over1+\beta^2/\betap^2}
={\betastar^2\over\betap^2+\beta^2},\quad\mbox{where}\quad
\betastar^2=\qpz\betap^2.
\label{qpbeta2}
\end{equation}
The resulting dependencies $\betap(\Rm)$, $\betastar(\Rm)$, and
$\qpz(\Rm)$ are shown in \Fig{pfitR}.
We see that $\betastar(\Rm)$ varies relatively little between 0.1 and 0.2 and is typically around 0.15.
For small values of $\Rm$, $\betap(\Rm)$ drops from 1 to 0.1 and stays then
approximately constant, while $\qpz(\Rm)$ rises proportional to $\Rm^2$
for $\Rm\leq10$ and then levels off at a value around 40.

The significance of a positive $\qs$ value comes from mean-field
simulations with $\qs>0$ indicating the formation of three-dimensional
(non-axisymmetric) flux concentrations \citep{BKR10}. This result was
later identified to be a direct consequence of having $\qs>0$ \citep{KBKR11}.
Before making any further conclusions, it is important to assess the
effect of other terms that have been neglected.
Two of them are related to the vertical stratification, i.e.\
additional terms in Eq.~(\ref{A1}) that are
proportional to $g_ig_j$ and $g_i\meanB_j+g_j\meanB_i$ with $\grav$
being gravity. The coefficient of the former
term seems to be small \citep{KBKMR11}
and the second only has an effect when there is a vertical
or inclined imposed magnetic field.
However, there could be other terms such as $\meanJ_i\meanJ_j$
as well as $\meanJ_i\meanB_j$ and $\meanJ_j\meanB_i$
when the scale separation is not enough large.

\section{Conclusions}

In the present paper we have
performed detailed investigations of NEMPI detected recently by BKKMR.
Most notably, we have extended the values of the scale separation
ratio, $\kf/k_1$, from 15 to 30.
In this case, the spontaneous formation of magnetic structures
becomes particularly evident and can clearly be noticed even without
any averaging.
Whether or not the particular structures seen in DNS
really have a correspondence to phenomena in the Sun,
cannot be answered at the moment,
because our model is still quite unrealistic in many respects.
For example in the Sun, $\kf$ and $\urms$ change with depth, which
is not currently taken into account in DNS.
Also, of course, the stratification is not isothermal, but convectively
unstable.
However, DNS in turbulent convection
by \cite{KBKMR11} have shown that $\Peff(\beta)$
still has a negative minimum in that case.

With regards to the production of sunspots, it is likely that
NEMPI will shut off before the magnetic energy density has reached values comparable with the
internal energy of the gas, as is the case in sunspots.
Thus, some other mechanism is still needed to push the field of flux
concentrations into that regime.
One likely candidate is the mechanism of KM where the suppression
of convective heat flux by the magnetic field is crucial.
This impression is further justified by resent calculations of \cite{Stein}
where pores are seen to form spontaneously in a simulation where horizontal
magnetic fields are injected at the bottom of the domain.

Pores are small sunspots, so something else is needed to make these
structures bigger and to amplify this mechanism further.
Again, the answer could be related to scale separation.
Thus, it will now be important to undertake detailed investigations
of instabilities in strongly stratified layers with finite heat flux and
finite magnetic field.

\begin{acks}
We acknowledge the NORDITA dynamo programs of 2009 and 2011 for
providing a stimulating scientific atmosphere.
Computing resources provided by the Swedish National Allocations Committee
at the Center for Parallel Computers at the Royal Institute of Technology in
Stockholm and the High Performance Computing Center North in Ume{\aa}.
This work was supported in part by the European Research Council
under the AstroDyn Research Project No.\ 227952.
\end{acks}

\end{article}
\end{document}